\documentclass{article}
\usepackage{graphicx} 
\usepackage[a4paper, total={5in, 10in}]{geometry}
\usepackage{multicol}
\usepackage{amsmath}
\usepackage[hidelinks]{hyperref}
\usepackage[citestyle=authoryear, sorting=nyt, uniquename=false]{biblatex}
\usepackage{fancyhdr}
\DeclareUnicodeCharacter{2032}{'}
\title{LSTM model predicting outcome of \\ strategic thinking task exhibits \\ representations of level-k thinking}
\author{\bf{Mario Stepanik}}
\date{Department of Mathematics, ETH Zürich\\mstepanik@ethz.ch}

\begin{document}

\maketitle

\begin{abstract}
Which neural mechanisms underlie strategic thinking in the human brain? Neuroeconomic research has not yet bridged the gap between theoretical models of higher-order reasoning and the precise mechanisms implemented in neural networks in the human brain. In this paper, I demonstrate that a recurrent neural network model can learn to perform strongly in the simple strategic game Rock-Paper-Scissors. In doing so, it develops implicit representations of strategically important variables (the levels $k$ of reasoning) which economists have postulated in theoretical models. These representations can be extracted from the hidden activations of the neural network. These findings hint at a connection between the mechanisms implicit in recurrent neural networks and models of strategic thinking in economic theory. Future empirical brain research can investigate whether these mechanisms correspond to mechanisms implicit in biological neural networks.
\end{abstract}


\section{Introduction}

For several decades, economists interested in strategic interaction have sought to develop theoretical models which describe and mechanistically explain human behavior in strategic situations. Early game theorists and mathematicians (\cite{nash1950equilibrium}, \cite{von1944theory}) centered upon equilibrium models in their analyses: these models characterize stable states in which no rational player has a profitable incentive to deviate from their chosen strategy, given the strategies of the other rational players (\cite{camerer2015psychological}). While such models have been useful in explaining equilibrium outcomes in a range of economic, political or other social situations, they have been criticized on grounds of the high cognitive requirements necessary for players to identify equilibrium strategies in general contexts. For example, to play in Nash equilibrium, the chosen strategies are required to be a set of mutual best responses: it is not enough for players simply to be rational, but they need to hold a \emph{correct} conjecture of what strategy other players are choosing, whereas rationality alone only requires players to react rationally to \emph{some} conjecture of what a rational player might do (\cite{osborne1994course}). In many relevant scenarios, fulfilling this requirement simply is not possible given cognitive constraints.

To remedy this shortcoming, behavioral game theorists developed a gamut of models to account for empirical findings from experiments investigating strategic behavior (\cite{camerer2010behavioural}). A key approach to achieve this aim is the formalization of players' cognitive limitations through level-$k$ models: first introduced by \cite{stahl1995players}, these models posit that players reason about their opponents' strategies by recursively assuming that their opponents are playing at a lower level of reasoning. These models have been applied in a wide range of contexts, including auctions (\cite{crawford2007level}), coordination games (\cite{nagel1995unraveling}), and signaling games (\cite{costa2008stated}).

Two main refinements of level-$k$ models – cognitive hierarchy and Bayesian level-$k$ models – have been developed to incorporate flexibility in reasoning about the opponent's strategy under uncertainty. Cognitive hierarchy models (\cite{camerer2004cognitive}) posit that players have a distribution of beliefs about the types of players they might encounter in a game, and they reason about their opponents' strategies based on these beliefs. Bayesian level-$k$ models (\cite{ho2021bayesian}) combine the recursive assumption of level-$k$ models with the distributional beliefs of cognitive hierarchy models. In Bayesian level-k models, players recursively update their beliefs about the levels of players to reason about their opponents' strategies.

A growing body of literature at the intersection of experimental psychology, neuroscience, and economics investigates the relationship between these theoretical models and the neural underpinnings of human behavior in strategic games. However, the mechanisms through which these feats of reasoning are accomplished in the brain have remained mostly elusive thus far. In this paper, behavioral data from a simple strategic thinking task (similar to Rock-Paper-Scissors) used in an experiment in our group\footnote{Laboratory of Prof. Christian Ruff, Department of Economics, University of Zurich} is used to train an artificial neural network (a long short-term memory [LSTM] model) with the aim of predicting future game outcomes. It is then investigated whether the hidden activations of the LSTM model represent variables relevant in the theoretical model, a Bayesian level-$k$ model developed in our group. The rest of this paper is organized as follows: in sections 2 and 3, the experimental set-up and the theoretical model used are described. In section 4 and 5, the methodology used is outlined and the results are presented. Section 6 offers a discussion of the key findings.

\section{Experimental set-up}

In the experiment conducted in our group, we investigated the neural mechanisms underlying strategic thinking using a simple paradigm akin to the simultaneous, zero-sum game Rock-Paper-Scissors (RPS). A total of 100 healthy participants were recruited and scanned using functional magnetic resonance imaging (fMRI) while playing 6 blocks of 40 rounds of RPS each.  They were instructed that their opponent was another human participant, but they in fact played against a bot programmed to take (partially noisy) choices corresponding to a level \emph{k} for $k\in\{0,1,2\}$ (described in detail below). Bar the noisy rounds, this level remained fixed over the course of one block. During the task, participants chose between actions $1$, $2$, and $3$ which correspond to rock, paper, and scissors. That is, playing $2$ when the opponent plays $1$ results in winning the round, playing $3$ in losing, and playing $2$ in a tie. These outcomes correspond to in-game rewards of 1, -1, and 0 points, respectively, whose sum determined the final monetary remuneration of participants.

In this task, a level-$0$ strategy is defined as a strategy in which each action $a\in\{1,2,3\}$ is chosen with a fixed probability $f_0(a)$. Note that any distribution is feasible, including uniform randomization across all possible actions which is the Nash equilibrium strategy. Thus, a level-$0$ player $j$ chooses action $a_j$ with probability $P(a_j|k_j=0)=f_0(a_j)$. 

The level-$1$ strategy is then defined as the strategy in which player $i$ chooses the best response to player $j$'s level-$0$ strategy. Thus, the action with the highest expected payoff given the (empirical) probability distribution $f_0(a_j)$ is chosen (in the case multiple best actions, randomization is possible). All strategies for levels $k>0$ are defined analogously as the best response to the strategy for level $k-1$. 

In this study, the bot played on each level $k\in\{0,1,2\}$ in two of the six blocks per player. Occasionally, the bot deviated from the action predicted by its level and chose a random action to avoid habituation effects. The frequency of this behavior depended on the human player's performance. Note that since the bot's strategy (i.e., level) remains unchanged over the course of a block, making it predictable to a substantial degree, it is not optimal to play the Nash equilibrium strategy. Attempting to predict the next action using information gathered over the previous rounds promises a higher success rate and thus, higher cumulative reward. This encourages participants to effectively play on level $k+1$ when the bot plays on level $k$.

\section{Theoretical model}

To formalize the reasoning process participants engage in when performing this task, our group developed a Bayesian level-$k$ model which predicts the player's posterior belief that the opponent is on a given level after observing some history of outcomes. 

First, let us define precisely what it means to play on some level $k$ in the context of the task described above. A level-$0$ player $j$ chooses action $a_j$ with probability $P(a_j|k_j=0)=f_0(a_j)$ and thus does not, by definition, update his strategy in the light of past outcomes. If $k_i>0$, player $i$ maximizes her expected payoff $\pi_i(a_i|\sigma_i)$ by playing action $a_i$ with probability $p_i(a_i|\pi_i;\sigma_i;k_i)$, where $\sigma_i$ represents the empirical estimation of the level-$0$ frequencies: in each round $t$, higher-level players ($k>0$) form beliefs about level-$0$ play according to the recent history $H_{\{t-T-1,t-1\}}$ of the game with memory length $T$. That is, player $i$ estimates the probability distribution $f_0$ of a level-$0$ player by calculating the frequency of each previous action. Due to limited memory, the frequency distribution is based on the last $T$ trials. Formally,

\begin{equation}
  \sigma_i(a_{0,t}|k_i)=
    \begin{cases}
      f_0(a_j|H_{\{t-T-1,t-1\}}) & \text{for $k_i\in\{1,3,5,...\}$}\\
      f_0(a_i|H_{\{t-T-1,t-1\}}) & \text{for $k_i\in\{2,4,6,...\}$}\\
    \end{cases}    
\end{equation}

Note that if player $i$’s level $k_i$ is odd (i.e., $k_i\in\{1,3,5,...\}$), then she assumes that the other player $j$ plays on level $0$, $2$, $4$, and so on. Therefore, player $i$ assumes that $j$ plays level $0$, or $i$ assumes that the other player $j$ assumes that $i$ assumes that $j$ plays on level $0$, and so on. The focus therefore lies on the history of the other player $j$ if player $i$ is playing an odd level $k$. Vice versa, if player $i$’s level $k$ is even, then the focus lies on the history of player $i$.

Second, let us define the action selection process. As described, if $k_i>1$, player $i$ maximizes her expected payoff $\pi_i(a_i|\sigma_i)$ by playing action $a_i$ with probability $p_i(a_i|\pi_i;\sigma_i;k_i)$. This probability is determined using the softmax function:

\begin{equation}
    p_i(a_i|\pi_i;\sigma_i;k_i) = \frac{\exp[m_i\cdot \pi_i(a_i|\sigma_i)]}{\sum\exp[m_i\cdot \pi_i(a_i|\sigma_i)]}
\end{equation}

Here, an inverse (and inherent) noise level of $m_i$ is used which reflects inherent randomization or mistake. We assume that the noise structure is based on the relative magnitudes of expected payoffs $\pi_i(a_i|\sigma_i)$. Also note that expected payoffs $\pi_i$ of player $i$ are based on the beliefs of player $i$ about the probability of actions played by player $j$ with $p_j(a_j)$. Specifically, player $i$ integrates over all beliefs and computes the expected payoff based on the distribution of player $j$’s actions:

\begin{equation}
    \pi_i(a_i)=\sum_{a_j}p_j(a_j)\cdot R_i(a_j|a_i)
\end{equation}

$R_i\in\{-1,0,1\}$ represents the outcome of the round with action pair $a_j$ and $a_i$.

Finally, in each round $t$, players with $k>0$ perform perfect Bayesian updating based on (i) their prior beliefs about levels and (ii) the actions observed in the given round $t$. At time step $t=0$, we have no information about the level on which the opponent plays. Thus, we use a uniform prior over levels $0$, $1$, $2$, and $3$ (note that the action chosen by a level-4 player is indistinguishable from the one chosen by a level-1 player – and so on – in this game). At every subsequent time step, the probability of an action $a_i$ originating from any level $k$ is given by $P(a_i|k)$ for $k\in\{0,1,2,3\}$. Using the probability distribution of action $a_i$ over all possible levels $k$, we can update our beliefs over all levels $k$ with $P(k)$ using Bayes' rule. Therefore, the posterior belief of any level $k$ after observing action $a_i$ is given by:

\begin{equation}
    P(k|a_i) = \frac{P(a_i|k)\cdot P(k)}{P(a_i)}
\end{equation}

\section{Methodology}

\subsection{Sequence prediction network}
\label{sec:seqpred}

\begin{figure}[t!]
  \centering
  \includegraphics[keepaspectratio, width=0.8\textwidth]{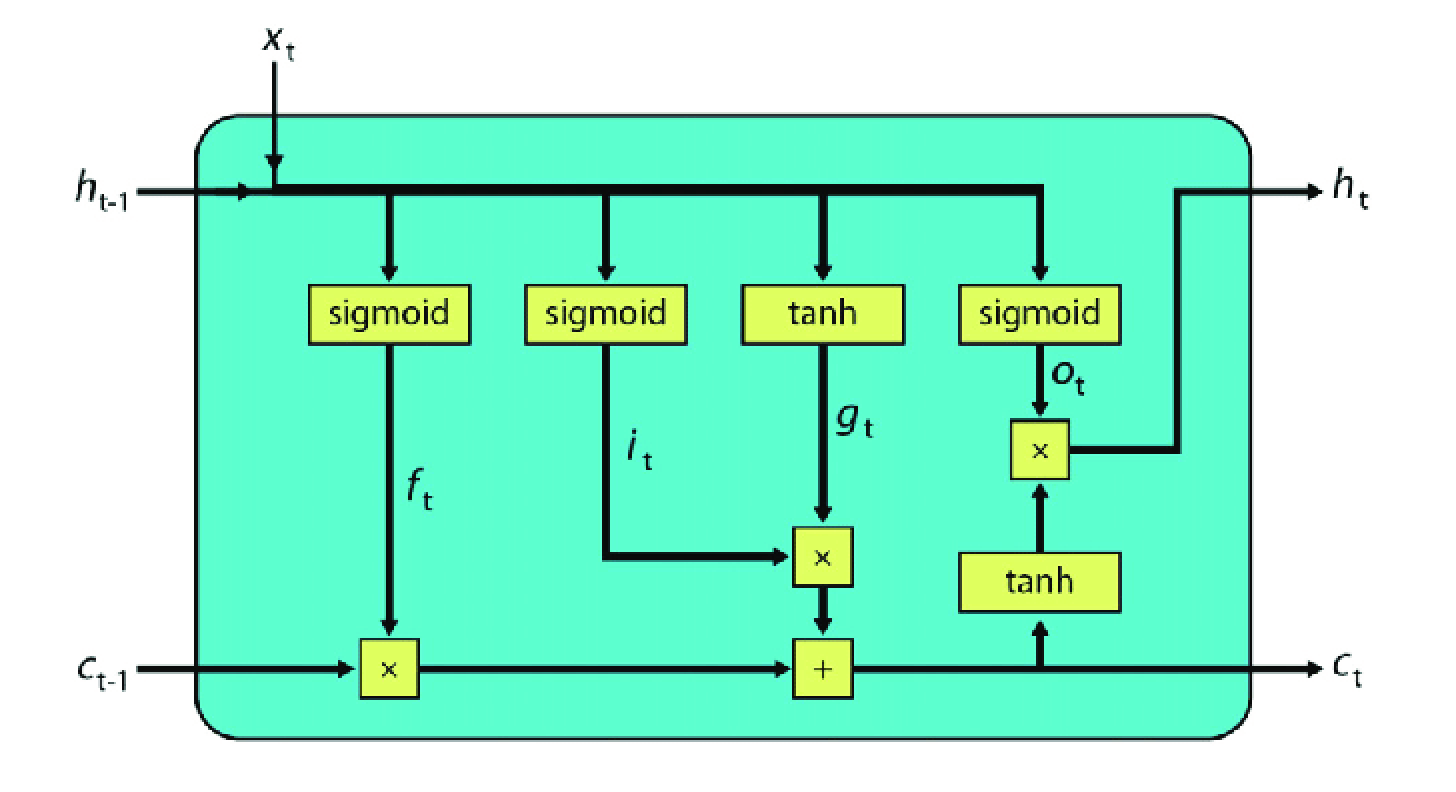}
  \caption{Computational graph of a LSTM cell. Illustration adopted from \cite{gokmen2018training}.}
  \label{fig:lstmgraph}
\end{figure}

The purpose of this paper is to investigate whether a neural network model which consists of long short-term memory (LSTM) cells (\cite{lstm}) develops level-like representations and employs an approximately Bayesian updating process when trained to predict sequences of RPS rounds. Note that the task of predicting the best next move given a history of round outcomes is essentially a sequence prediction problem. LSTM models are particularly well suited for this problem: for over twenty years, neural network architectures using LSTM layers have been a major building block in constructing sequence prediction models as (i) they do not suffer from the vanishing gradient problem as traditional recurrent neural networks do and (ii) they can learn important relationships between data points even when they are distant in time (\cite{lstm}).

\begin{figure}[t!]
  \centering
  \includegraphics[keepaspectratio, width=1\textwidth]{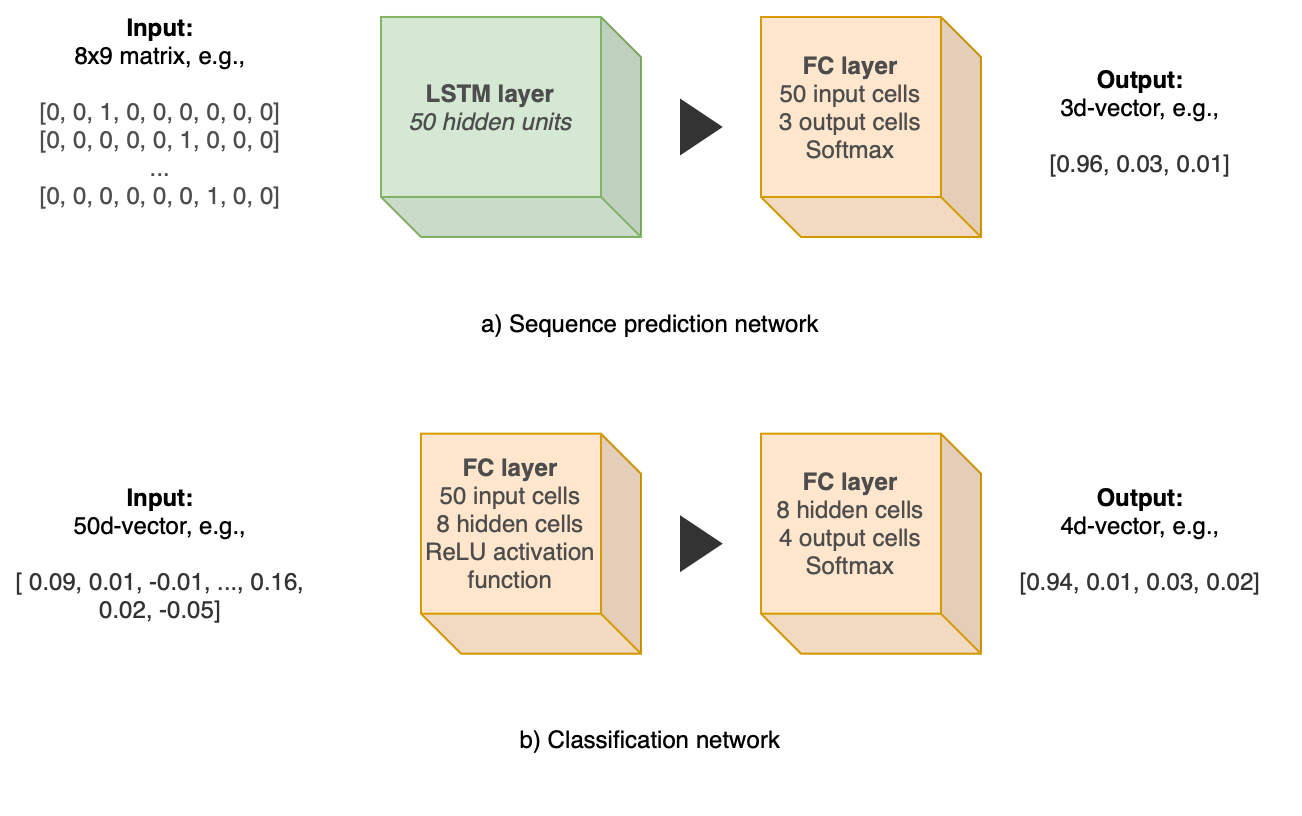}
  \caption{Neural network architectures used in the simulations. FC stands for a "fully connected". \textbf{a)} Sequence prediction network structure to predict next game outcome. \textbf{b)} Classification network structure to extract bot level.}
  \label{fig:network}
\end{figure}

The key problem the human participant faces is the correct prediction of the opponent's next action, for if she succeeds in doing so, winning the round is trivial. Hence, this was the training objective of the present model (although predicting the human's next action and the overall round outcome was successful as well; see \hyperref[sec:results]{Results}). Each round outcome was encoded as a nine-dimensional one-hot vector, representing the $3^2$ possible game outcomes in RPS. The data was segmented into series of 36 rounds as some rounds were missing in several blocks. It was found that the optimal sequence length to be used as training input was 8 (i.e., one training sample consists of an $8\times 9$ matrix which contains the concatenated one-hot vectors representing time step $t$ to $t+7$, the target being the one-hot vector at time step $t+8$). These sequences were fed into a PyTorch LSTM layer (\cite{pytorch}) with 50 hidden units, which effectively implements the following calculations (see also Figure ~\ref{fig:lstmgraph}):

\begin{align*}
    i_t &= \sigma(W_{ii}x_t+b_{ii}+W_{hi}h_{t-1}+b_{hi})\\
    f_t &= \sigma(W_{if}x_t+b_{if}+W_{hf}h_{t-1}+b_{hf})\\
    g_t &= \tanh(W_{ig}x_t+b_{ig}+W_{hg}h_{t-1}+b_{hg})\\
    o_t &= \sigma(W_{io}x_t+b_{io}+W_{ho}h_{t-1}+b_{ho})\\
    c_t &= f_t \odot c_{t-1} + i_t \odot g_t\\
    h_t &= o_t \odot \tanh(c_t)
\end{align*}

where $h_t$ is the ($50\times 1$)-hidden state at time $t$, $c_t$ is the ($50\times 1$)-cell state at time $t$, $x_t$ is the ($8\times 9$)-input at time $t$, $h_{t-1}$ is the ($50\times 1$)-hidden state of the layer at time $t-1$ or the initial hidden state at time $0$, and $i_t$, $f_t$, $g_t$, and $o_t$ are the input, forget, cell, and output gates, respectively (\cite{pytorchlstm}). An LSTM cell tracks the cell and hidden states, whose activations are updated at each time step in the light of the new input. During the training phase, the input weights ($W_{ik}$ for $k\in\{i,f,g,o\}$) and biases ($b_{ik}$) as well as the hidden weights ($W_{hk}$) and biases ($b_{hk}$) are updated via stochastic gradient descent with batch size 16 (using Adam optimizer) and a learning rate of $0.01$ to minimize cross-entropy loss. The final hidden states constitute the output of the LSTM layer, which is fed into a fully connected layer with three output units corresponding to the three actions. Finally, the softmax function is applied to obtain prediction probabilities. This architecture is illustrated in Figure ~\ref{fig:network}a). The hyperparameters (number of hidden units, sequence length, batch size, learning rate, etc.) were chosen to minimize loss on the test set (see the code accompanying the paper for details).

\subsection{Level classification network}

To answer the question of whether the LSTM network learns to represent the level on which the bot plays in a given round in order to solve to prediction task, I hypothesized that the hidden activations $h_t$ at a given time step $t$ would need to contain increasingly useful information for predicting the level over the course of one round. That is, at time step $t=0$, we would expect no meaningful information to be encoded in the hidden activations whereas we could extract the level at later time steps. To test this hypothesis, a separate minimal classification network with only one hidden layer was trained for every time step $t\in[0,35[$. The 50 hidden activations given the input at time $t$ were used as the training features and the bot's ground-truth level as its target. The separation into different networks for each time step is crucial as it enables us to evaluate the confidence the network has in its prediction regarding the level. For example, at time step $0$, providing the ground truth label should be relatively uninformative, which would be reflected in level predictions around the uniform prior over all considered levels. At time step $35$, in contrast, we would expect not only a correct prediction of the level on which the bot plays, but also a confidence of close to 1 that this prediction is correct, as the Bayesian level-$k$ model predicts. The architecture used in this network is illustrated in Figure ~\ref{fig:network}b.

\section{Results}
\label{sec:results}

\subsubsection*{LSTM network predicts next action with high accuracy}

After training, the LSTM network architecture depicted in Figure ~\ref{fig:network}a) achieves a test-set accuracy of 92$\%$ in predicting the next action of the bot (excluding the noisy rounds which cannot be predicted from the accuracy calculation). Chance level accuracy for this task is 33$\%$. Thus, the model evidently learns a meaningful representation of the sequence structure. Decreasing the number of hidden units leads to a fall-off in accuracy while increasing it has no significant effect. Neither did adding additional LSTM or fully connected layers.

Additionally, the model also performs reasonably well in predicting the human player's actions, in which it achieves an accuracy of 59$\%$ (chance level: 33$\%$), and in predicting the overall outcome of the round, in which it achieves an accuracy of 44$\%$ (chance level: 11$\%$). It is rather unsurprising that the model's performance is significantly stronger in the case of the bot as its actions are entirely deterministic given the level (bar the noisy rounds). 

These results can be reproduced by running the accompanying code.

\subsubsection*{Level of play can be decoded from network activations with high accuracy}

\begin{figure}[t!]
  \centering
  \includegraphics[keepaspectratio, width=0.8\textwidth]{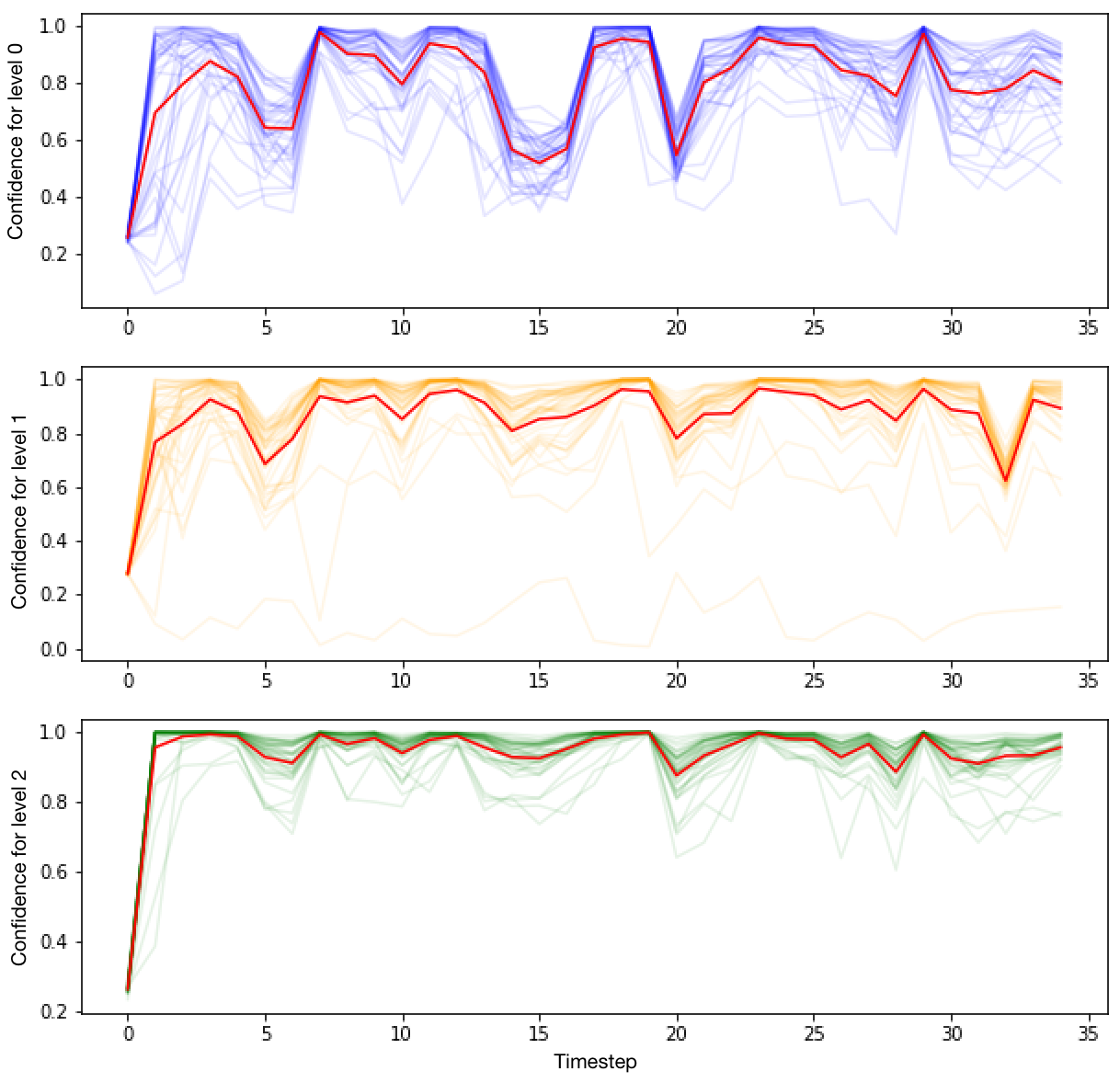}
  \caption{Confidence of the classification network that the bot plays on level $k\in\{0,1,2\}$. The fine lines represent the predictions for $n=120$ test-set blocks. The red line represents their mean. \emph{Top}: $n_0=40$ cases in which the ground-truth level of the bot was 0. The confidence increases from the uniform prior's prediction of 0.25 at $t=0$ to a high confidence swiftly. \emph{Middle}: $n_1=40$ cases in which the ground-truth level of the bot was 1. \emph{Bottom}: $n_2=40$ cases in which the ground-truth level of the bot was 2.}
  \label{fig:levelacc}
\end{figure}

\begin{figure}[t!]
  \centering
  \includegraphics[keepaspectratio, width=1\textwidth]{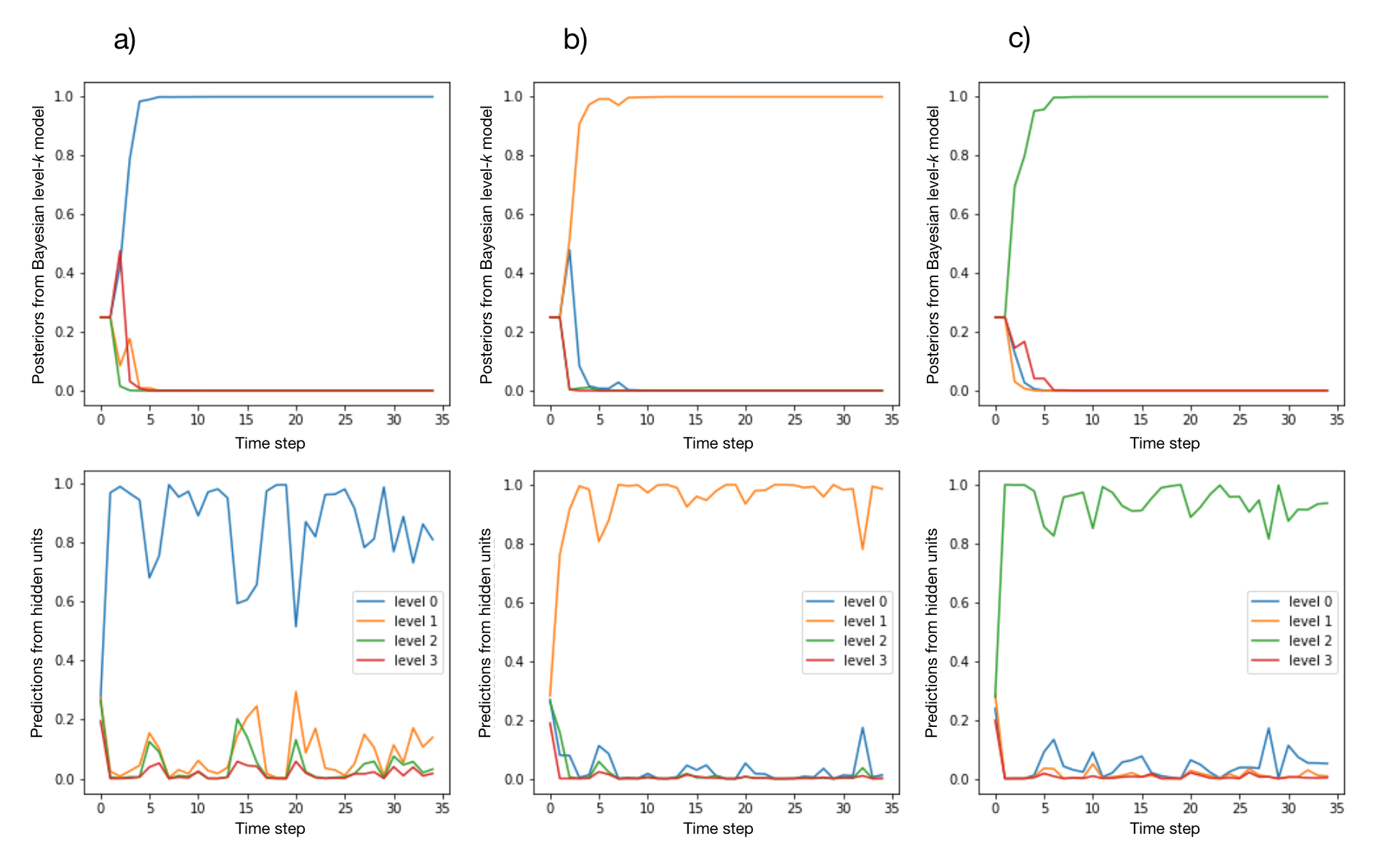}
  \caption{Examples of the change of the Bayesian posteriors for different levels over the course of one block of rounds according to the theoretical model (\emph{top row}) and the predictions of the classification network based on hidden units for individual blocks from the test set (\emph{bottom row}). \textbf{a)}: Example of convergence when the bot plays on level 0. \textbf{b)}: Example of convergence when the bot plays on level 1. \textbf{c)}: Example of convergence when the bot plays on level 2.}
  \label{fig:postlstm}
\end{figure}

As hypothesized, it was found to be the case that a minimal classification network can extract an increasing amount of information about the level on which the bot plays from the hidden activations $h_t$ at time step $t$ generated by an input sequence. In fact, we can observe that the confidence in identifying the level increases in a similar fashion to how the theoretical Bayesian level-$k$ predicts. Figure ~\ref{fig:levelacc} illustrates the prediction confidences of the classification network for levels $0$, $1$, and $2$ for those rounds in which these are the respective ground-truth levels (note that the network was trained to distinguish level $3$ as well, but this level was never actually played by the bot). As predicted, the network cannot extract meaningful information about the level at time step $0$, hence its average prediction confidence is close to 0.25, which corresponds to the uniform prior. Within a few time steps, the model achieves a high confidence in its predictions and is correct in over $99\%$ of test cases. In Figure ~\ref{fig:postlstm}, the posterior beliefs predicted by the Bayesian model and the predictions made by the classification network are depicted for three individual blocks chosen from the test set for illustration.

We can observe that the network is more confident in its predictions when the bot plays on level $1$ or $2$ compared to $0$. It stands to reason that this is a consequence of the level definitions in the theoretical model: while the next actions of all levels $k>0$ are deterministically defined as best responses to level $k-1$ (bar randomization in the case of ambiguity), a level-$0$ player chooses his action $a$ randomly according to a probability distribution $f_0(a)$. Hence, there is an inherent difficulty in distinguishing between (random) level-$0$ play and (essentially deterministic) level-$k$ play (for $k>0$), which makes it more challenging to confidently identify level-$0$ play.

\begin{figure}[t!]
  \centering
  \includegraphics[keepaspectratio, width=1\textwidth]{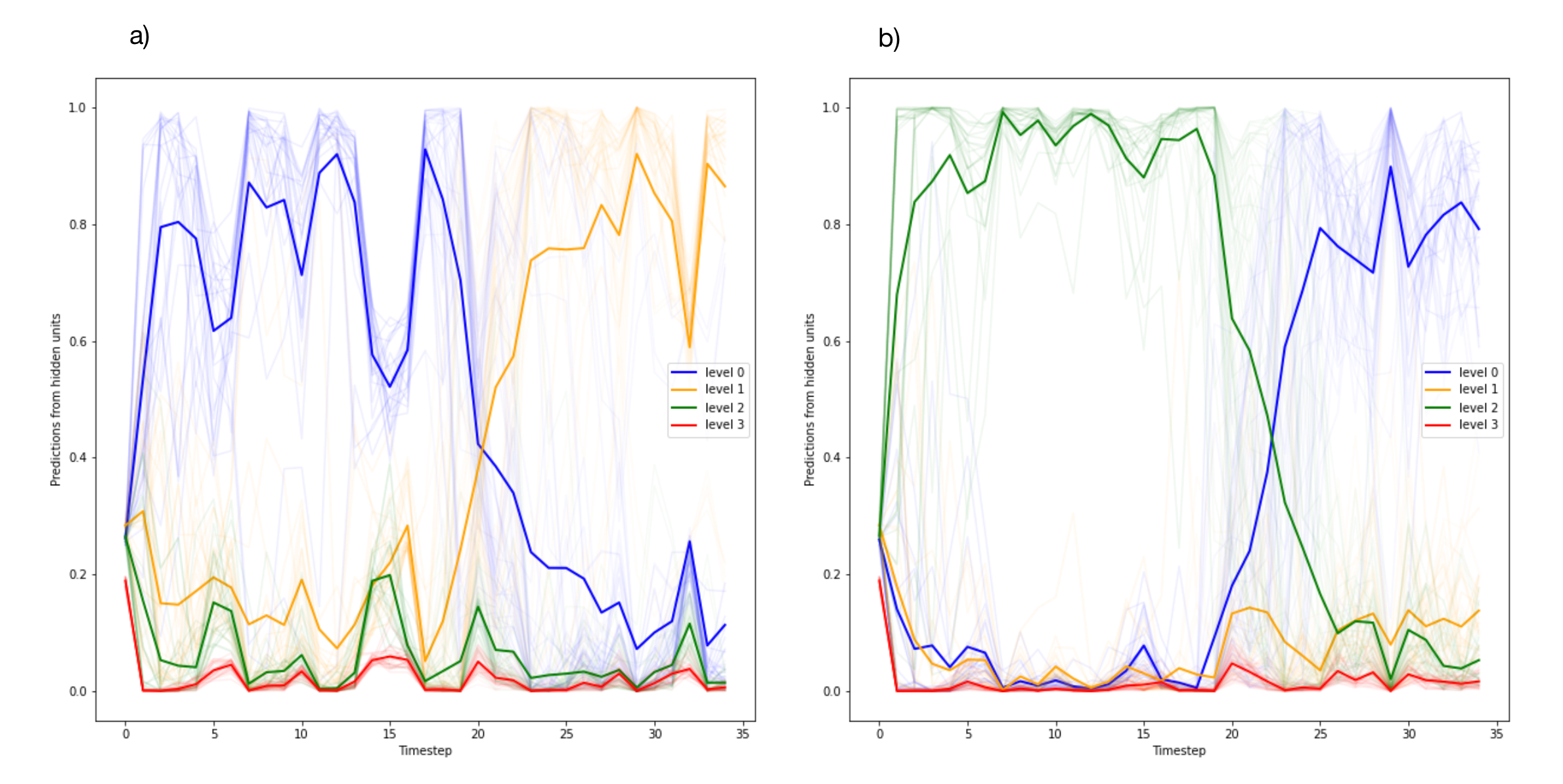}
  \caption{Level of bot predicted by the classification network from hidden activations $h_t$ for concatenated rounds with a level change. The level change happens at $t=18$. Depicted are the means of all test-set level changes \textbf{a)} from level 0 to level 1 and \textbf{b)} from level 2 to level 0. Other transitions can be plotted with the accompanying code.}
  \label{fig:leveltrans}
\end{figure}

To corroborate the hypothesis that the hidden activations $h_t$ dynamically encode levels of play, I ran another set of counterfactual experiments. Rather than using the original round sequence with constant level play of the bot, I split up and concatenated two different sequences of the same subject in which the bot plays on different levels. That is, the round outcomes for time steps $t=0,...,17$ were taken from block $b_1$ of subject $s$ in which the bot played on level $k_1$, while for the outcomes for $t=18,...,34$ were taken from block $b_2$ of subject $s$ in which the bot played on level $k_2$.\footnote{Of course, this counterfactual experiment suffers from the limitation that if the bot had in fact changed its level of play, the human opponent may have reacted differently in the first rounds after the change. However, this should only make it easier to identify a level change.} The resulting prediction confidences for different $k_1-k_2$ transitions are depicted in Figure ~\ref{fig:leveltrans}. We can indeed see that after a short period of uncertainty directly after the level change, the network confidently identifies the new level. Together, these experiments substantiate the claim that the LSTM model develops a similar representation of high-level strategic variables to solve the task as postulated by economic theory.

Finally, note also that not only hidden states $h_t$, but also cell states $c_t$ as well as the activations of the internal gates ($i_t$, $f_t$, $g_t$, and $o_t$) allow for such decoding of levels. This is rather unsurprising as these activations are, according to the definition of the LSTM cell, simply the result of nonlinear transformations of the hidden states $h_t$ (see \hyperref[sec:seqpred]{Sequence prediction network} for the corresponding formulae). Hence, it seems reasonable that information about level play is also encoded in these activations.

\subsubsection*{Individual activations are not easily interpretable in terms of strategic variables}

Visual inspection of the individual activations $h_{ti}$ for $i\in{1,...,50}$ reveals no obvious order which corresponds to variables or elements one would expect to be relevant to solve the task. This indicates that the information about levels which can be extracted by the classification network described above is not encoded in individual activations, but in their (potentially nonlinear) combination.\footnote{The individual hidden activations can be plotted with the accompanying code.} Future research could investigate multiple intriguing questions in this regard to move closer towards a mechanistic understanding of LSTM networks: what role do the calculations performed by individual gates play in creating the emergent properties such as level representation? Can the contributions of individual units be disentangled to identify individual tasks such as enabling translation invariance (i.e., which units enable steady performance if action labels are rotated) versus level representation? Answering these questions could be an alternative, more theory-driven approach to interpret the functioning of LSTM networks (\cite{pmlr-v97-guo19b}, \cite{krakovna}).

\section{Discussion}

The results presented here indicate that a LSTM network acquires hidden representations necessary for level-$k$ thinking. In the light of the multitude of possible approaches towards solving sequence prediction tasks (such as Markov models or directed graphs), it is noteworthy that a neural-network-based approach seems to represent the core elements of the theoretical level-$k$ model conceived by behavioral game theorists. This suggests that the theoretical economic representation of game play may indeed be an efficient approach towards solving the task.

There are two key questions which deserve consideration when we assess the correspondence between the mechanisms employed by the LSTM model and the theoretical model. 

First, we have not yet ascertained that the distinguishable representations encoded in the hidden activations $h_t$ in fact constitute \emph{levels}. It is conceivable that the LSTM model identifies four different algorithms, each of which corresponds to a level $k$ in the theoretical model, but that these are simply distinct but unordered strategies. The key characteristic of level-$k$ thinking is that higher levels correspond to higher orders of thinking. That is, level $k+2$ is in some sense "further" from level $k$ than level $k+1$. Most likely, it is not possible to establish whether this is represented in the model in such a simple task. Note that the player only has three actions to choose from and that higher levels ($k>3$) are largely indistinguishable from lower levels solely from behavior.\footnote{In other tasks in which levels are more clearly distinguishable, one could for example permute the labels in the training data of the classification network to see whether this violation of the level structure has a detrimental impact on prediction performance.} 

Second, it is unclear whether the updating of hidden activations $h_t$ can be called "Bayesian". Bayesian updating is characterized by the multiplication of a prior and a likelihood with subsequent normalization. While we can observe an updating process of "beliefs" in the light of new inputs, there is no evidence that this updating is done via Bayes' rule. On the one hand, one could argue that the successful application of a principle (in this case, Bayesian updating) does not require accurately performing the calculations prescribed by theory, but merely an approximation which yields a sufficiently similar result (think of the exemplum of the expert pool player whose skill is the consequence of extensive training, but whose behavior can be very well approximated by and predicted with the laws of mechanics, even if he is not aware them). On the other hand, it is well possible that Bayesian updating is simply not a good description of the model's behavior. The counterfactual experiments described above indicate that this is the case: while the LSTM model can adapt to a change in level flexibly within a small number of time steps, the Bayesian model virtually reaches deadlock once the posterior probability of a level has converged close to 1 (a phenomenon called the "Bayesian trap"). If this happens, a large amount of evidence is necessary to move away from this prediction. This is clearly not a problem the LSTM model suffers from. I would characterize the LSTM behavior as that of an educated guesser: upon observing an action which is in accord with play on some set of levels (e.g., level 0 or level 2, which are not distinguishable after observing only one round which indicates level 2 play), it assumes with high confidence that the bot is on one of those levels and revises this belief if contradictory evidence comes in. In fact, one can observe that in some test cases (not depicted), the network predicts level-2 play very confidently after only one time step, but revises its belief to level-0 play (which is the correct level) later on.

The findings presented here open up a set of intriguing questions for neuroscientists and economists. While a number of brain areas implicated in level-$k$ thinking have been identified experimentally (\cite{coricelli2009neural}, \cite{lee2016neural}, \cite{saxe2003}), it is still unclear precisely which mechanisms the brain would employ to implement level-$k$ thinking, assuming that it does. The present simulation study demonstrates that a LSTM architecture is able to solve the given strategic thinking task and that the model's activations encode variables which are thought to be theoretically necessary for achieving this. This prompts the question of whether the brain may use a neural network architecture similar to a LSTM model. Future research could investigate whether there is a correspondence between the activation of hidden units, cell units or internal gates in the artificial LSTM network and brain activations measurable with brain scanning methods. This study shows that the calculations necessary to perform well in the strategic thinking task can be performed with an architecture based on artificial neurons as the fundamental computation unit. Future studies could build on these results to substantiate or disprove the hypothesis that the human brain uses a LSTM-like biological network architecture to solve the same task.

\section{Conclusions}

This paper demonstrated that an artificial recurrent neural network is able to perform well in a simple strategic thinking task and implicitly learn representations of variables postulated to be important by economic theory. In particular, it substantiates the claim that level-$k$ thinking is an efficient approach to solving this kind of task. However, the evidence suggests that the process of updating beliefs about the opponent's level of play in the light of new evidence from game outcomes is not Bayesian in the sense that Bayes' rule is applied. Nonetheless, the LSTM architecture used here strongly performs in the strategic thinking task and future research may investigate to which degree the mechanisms of the LSTM cell could also be implemented in brain structures.

\section*{Supplementary material}

The code used to produce the results and plots can be found and run \href{https://colab.research.google.com/drive/1nhMQ4wY0ZFa_goAiO800c7Zbacn9zKSV?usp=sharing}{\underline{here}}. The raw data document \texttt{data\_subs.xlsx} can be downloaded \href{https://docs.google.com/spreadsheets/d/1gEqj3zuVkOUBdCXwNYM1ANQIu0vMEW9e/edit?usp=sharing&ouid=102383954653729091628&rtpof=true&sd=true}{\underline{here}} (please request access).

\section*{Acknowledgements}

Thank you to my supervisors Prof. Dr. Christian Ruff, Prof. Dr. Martin Mächler, and Dr. Gökhan Aydogan for providing many useful comments, insights, and new directions for thought.

\printbibliography

\end{document}